# TLSCheck 2.0: An Enhanced Memory Forensics Approach to Efficiently Detect TLS Callbacks

*Kartik N. Iyer[1,\*] and Parag H. Rughani[2]*
[1]*Independent Researcher, Nagpur, Maharashtra, India*
[2]*National Forensic Sciences University, Gandhinagar, Gujarat, India*

**Abstract**

Memory analysis is a crucial technique in digital forensics that enables investigators to examine the runtime state of a system through physical memory dumps. While significant advances have been made in memory forensics, the detection and analysis of Thread Local Storage (TLS) callbacks remain challenging due to their dual nature as both legitimate Windows constructs and potential vectors for malware execution. An early version of the *TlsCheck* plugin received recognition in the Volatility Plugin Contest 2024. In this paper, we present an enhanced version of *TlsCheck* for Volatility 3, designed to detect and analyze TLS callbacks in process memory. It implements precise detection of TLS callback tables through analysis of PE headers and memory structures, combined with disassembly of identified callback routines. The plugin supports both 32-bit and 64-bit architectures, offering investigators insights into callback locations, assembly behavior, and potential signs of suspicious activity. To enhance detection, we incorporate pattern matching using custom regular expressions and YARA rules, helping analysts identify specific code patterns or suspicious constructs within TLS callbacks. The framework also includes instruction-level analysis to highlight behavior often linked to malware, such as anti-debugging, code injection, and process manipulation. This implementation significantly improves defenders' ability to detect and investigate TLS-based threats during memory forensics, supporting more effective malware analysis and incident response operations.

## 1. Introduction

Memory forensics has become an essential branch of digital forensics, providing the means to recover transient evidence that would otherwise be lost using traditional techniques, which often involve shutting down systems and analyzing non-volatile storage [1]. This transient data includes critical information such as active processes, network connections, encryption keys, and fragments of volatile application data. The importance of memory forensics has increased, especially as modern malware tends to avoid leaving traces on storage devices and instead uses memory-resident techniques.

The field gained significant momentum in 2005 when the Digital Forensic Research Workshop (DFRWS) launched the Memory Analysis Forensics Challenge [2]. This initiative propelled the development of specialized tools for analyzing memory dumps, including MoonSols, KntTools, FATKit, VolaTools, and the now widely used Volatility framework [4]. These tools have enabled investigators to extract actionable insights from volatile memory, proving indispensable in cases involving advanced cyber threats.

The significance of memory forensics was underscored in 2021 during two major cyberattacks: the SolarWinds supply chain attack and the exploitation of vulnerabilities in Microsoft Exchange [3]. Both incidents heavily relied on memory-only payloads, highlighting the increasing sophistication of malware that leverages memory to evade detection.

In this context, understanding and detecting specific memory-based techniques, such as Thread Local Storage (TLS) callbacks and their injection mechanisms, has become a critical focus for researchers and practitioners. TLS callbacks, which allow code execution at specific points during process initialization, present challenges and opportunities for forensic investigation. Their potential misuse by malware has been noted by MITRE Framework [5], emphasizing the need for robust detection and analysis tools to effectively identify and dissect these artifacts.

Our research advances memory forensics by introducing a specialized plugin for detecting and analyzing Thread Local Storage (TLS) callbacks – TlsCheck, in Windows processes, developed for the latest version of the Volatility framework, Volatility 3. TLS callbacks represent a niche and underexplored area in memory forensics, despite their significant role in process initialization. These callbacks execute code at specific points during a process's lifecycle, making them critical yet often overlooked artifacts in forensic investigations. Our plugin enables forensic practitioners to identify and disassemble TLS callbacks, examining their instructions for suspicious patterns that may indicate suspicious behavior. By addressing this specialized area, our work enhances the capabilities of memory forensics and provides a valuable tool for investigators, which will be made available to the open-source community for further exploration.

The remainder of this paper is organized as follows. Section 2 discusses related work in the relevant field. Section 3 identifies the gaps in current research and provides the motivation for this study. Section 4 outlines the contributions of this paper. Section 5 introduces the fundamental concepts necessary for understanding our work. Section 6 details the methodology employed. Section 7 addresses the limitations of the study and suggests directions for future research. Finally, Section 8 concludes the paper.

## 2. Related Work

Since the first release of Volatility Framework in 2007 [6], it has grown significantly, thanks to the active contributions of its community. Over the years, numerous plugins and extensions have been developed and continuously improved. The Volatility Plugin Contest has played a vital role in fostering innovation and collaboration, encouraging participants to push the framework's boundaries and enhance its capabilities. The contest has showcased a wide range of creative entries, resulting in plugins designed to identify, detect, and analyze various components within memory.

There have been limited studies specifically focused on analyzing TLS callbacks through memory forensics, despite existing evidence highlighting how TLS callbacks can be exploited for malicious purposes. Previous advancements in memory forensics have significantly enhanced callback analysis, exemplified by Volatility's callbacks plugin [7]. This plugin is designed to identify various types of kernel callbacks, such as filesystem registration changes, shutdown notifications, generic kernel callbacks, and bugcheck-related events. It achieves this through advanced pool scanning techniques and in- depth analysis of kernel structures, making it an effective tool for detecting callback mechanisms that malware could exploit for persistence or system monitoring. Similarly, tools like the malfind plugin [9] in both Volatility and Rekall frameworks provide powerful capabilities for detecting potentially malicious code by analyzing Virtual Address Descriptors (VADs) [8]. The malfind plugin focuses on VADs with WRITE and EXECUTE permissions, which are often indicative of code injection, offering a critical means of identifying malicious activity within memory.

## 3. Gaps and motivation

Despite the advancements discussed previously, notable gaps remain in the field of memory forensics, particularly regarding the analysis of Thread Local Storage (TLS) callbacks. While the callbacks plugin is adept at detecting kernel-level callback mechanisms, it does not address TLS callbacks, which malware increasingly leverages for execution and persistence. Similarly, the malfind plugin, although effective for identifying injected code, is not equipped to analyze the specific challenges posed by TLS callbacks. Additionally, though this aspect is extremely trivial, the malfind plugin has a fixed disassembly output length, limiting the amount of code visible to the user. To view more disassembly output, users are required to switch to the volshell for manual inspection, which can be time- consuming and disrupt the workflow. TlsCheck addresses this limitation through the --*disasm-bytes* flag (discussed in Section 6.2), allowing users to customize the number of bytes disassembled for each callback.

A notable example highlighting the significance of TLS callbacks is the paper titled How Malware Defends Itself Using TLS Callback Functions [10]. This article examines how malware, such as the Nadnazz bot, uses TLS callbacks to evade detection. For instance, when the bot's executable is loaded into OllyDbg, the debugger does not pause at the expected entry point, illustrating the stealth capabilities enabled by TLS callbacks. Another case is the Ursnif malware (2017) [11] [12], which employs TLS callbacks to obscure its execution flow and resist detection. These examples emphasize the critical need for dedicated tools to analyze TLS callbacks in memory forensics, addressing a significant gap in existing methodologies and providing essential insights into advanced malware behaviors.

## 4. Contributions

An initial version of *TlsCheck* has already received recognition in the Volatility 3 Plugin Contest 2024 [16]. In this paper, we present an enhanced version of *TlsCheck*, a novel Volatility 3 plugin that significantly enhances the analysis of Thread Local Storage (TLS) callbacks during memory forensics. TLS callbacks are often overlooked execution points within processes, yet they can be leveraged for both legitimate purposes and stealthy malicious behavior. Our work goes beyond simply extracting TLS callback addresses; *TlsCheck* introduces context-aware analysis by automatically disassembling and scanning these callbacks using multiple detection heuristics. These heuristics evaluate both syntactic patterns and semantic context to identify potentially suspicious behaviors such as NOP sleds, API hashing, control flow anomalies, etc (see Section 5.3). By combining extraction with intelligent analysis, TlsCheck transforms TLS callback examination from a traditionally manual, multi-step process into a streamlined, automated investigative capability that exposes sophisticated evasion techniques hiding in these overlooked execution points.

## 5. Fundamentals

Before diving into how our plugin works, it's necessary to first understand a few key concepts. This section will cover important ideas that are essential for effectively using and getting the most out of the plugin.

*5.1. Understanding processes and threads*

A process is the fundamental unit of execution in a computer system, providing the resources necessary to run a program. It comprises a virtual address space, executable code, open handles to system objects, a security context, environment variables, and a unique process identifier [13]. Additionally, processes are allocated a priority class and defined working set sizes, which determine the minimum and maximum memory available for execution. A process always includes at least one thread of execution, referred to as the primary thread, which is responsible for initiating the process's operations. However, processes can create additional threads to handle concurrent tasks or enhance performance [13].

Our plugin, TlsCheck, focuses specifically on extracting Thread Local Storage (TLS) callbacks for individual processes in a memory dump. TLS callbacks are specialized functions that execute during specific stages of a process's lifecycle, such as initialization or termination. Identifying these callbacks is critical for understanding how processes are configured and operate in memory, particularly in debugging or reverse engineering scenarios.

Within a process, threads are the entities that carry out the actual execution of code. All threads within a process share the same virtual address space and system resources, enabling efficient inter-thread communication. However, each thread maintains its own unique context,

```
(layer_name) >>> dt("_IMAGE_DATA_DIRECTORY")
symbol_table_name1!_IMAGE_DATA_DIRECTORY (8 bytes)
0x0   : VirtualAddress     symbol_table_name1!unsigned long
0x4   : Size               symbol_table_name1!unsigned long
```

**Fig. 1.** Representation of the structure IMAGE_DATA_DIRECTORY using Volshell.

```
typedef struct _IMAGE_TLS_DIRECTORY {
    DWORD   StartAddressOfRawData;  // Pointer to the beginning of TLS raw data
    DWORD   EndAddressOfRawData;    // Pointer to the end of TLS raw data
    DWORD   AddressOfIndex;         // Pointer to a variable holding the TLS index
    DWORD   AddressOfCallBacks;     // Pointer to an array of TLS callback function pointers
    DWORD   SizeOfZeroFill;         // Size of zero-initialized data for TLS
    DWORD   Characteristics;        // Reserved; typically set to zero
} IMAGE_TLS_DIRECTORY32, *PIMAGE_TLS_DIRECTORY32;
```

**Fig. 2.** Structure of IMAGE_TLS_DIRECTORY in Windows PE Format

also includes the thread's machine registers, kernel stack, user stack, and a thread environment block, all of which the system uses to manage and execute the thread. Threads can even have their own security contexts, allowing them to perform operations on behalf of different users or clients when required.

*5.2. Introduction to Thread Local Storage (TLS)*

Threads within a process share the same virtual address space, which enables seamless communication and resource sharing. However, the scope of data within this space varies. Local variables of a function are unique to each thread running the function, ensuring that each thread has its own independent execution context. On the other hand, static and global variables are shared across all threads in the process, which can lead to potential data conflicts if not carefully managed. To address this, TLS provides a mechanism to maintain thread-specific data that can be accessed globally within the process. As described in the article written by Karl Bridge et al. [14], this allows each thread to have its own unique data, retrievable through a globally allocated index.

In a Windows executable, the Thread Local Storage (TLS) directory is located in the Optional Header of the PE (Portable Executable) format, which is itself part of the NT Headers. Within the Optional Header, there is a Data Directory array, which contains entries pointing to various sections of the executable. The TLS directory is represented by the 10th entry in this Data Directory array. This entry contains the virtual address and size of the TLS directory structure (see figure 1). The TLS directory, defined as IMAGE_TLS_DIRECTORY, includes several key fields that describe the TLS implementation for the process (see figure 2). Among these fields is AddressOfCallbacks, which points to an array of function pointers. These functions, known as TLS callbacks, are executed automatically by the system during specific events in a process's lifecycle, such as initialization or termination. Other fields in the TLS directory provide information about the starting address of the TLS data, its size, and alignment requirements.

Our plugin leverages this architecture to extract TLS callbacks from IMAGE_TLS_DIRECTORY structure of an executable. By analyzing this structure, the plugin including exception handlers, scheduling priority, thread-local storage, and a distinct thread identifier. This context identifies and extracts the TLS callbacks associated with a process. This information is invaluable for understanding process initialization, resource management, and behavior in memory, particularly in scenarios such as debugging or reverse engineering.

*5.3. Attack Vectors in TLS Callback Mechanisms*

The 2017 Ursnif malware [12] was a variant that leveraged TLS callbacks to evade detection by executing malicious code during the callback phase. Inspired by this technique, we designed our plugin to incorporate a detection mechanism for commonly used suspicious instructions observed in the wild.

*5.3.1. Code Injection via Dynamic Memory Access*

This attack involves suspicious dynamic memory operations where attackers manipulate memory addresses using register combinations to inject suspicious code. The attack bypasses standard memory protections by using legitimate-looking memory access patterns but actually writes executable code to arbitrary memory locations. This is particularly dangerous in TLS callbacks since they execute before the main program entry point.

*5.3.2. Control Flow Hijacking*

This attack occurs when malware directly manipulates program execution flow through register-based calls or jumps, such as `call eax` or `jmp ebx`. While these jumps are common in legitimate code, their presence in an unknown context is suspicious, as they can be used for indirect calling or code obfuscation attempts. Within TLS callbacks, attackers can redirect program execution to suspicious code before normal program initialization. The code specifically looks for suspicious patterns where control flow is modified without proper context or legitimate function call structures.

*5.3.3. Stack String Constructions*

Stack string construction is a technique where attackers build strings character by character on the stack to hide potentially suspicious strings from static analysis. The code detects this through multiple patterns: (1) consecutive push operations with immediate values containing printable characters, (2) byte-by-byte construction using sequential mov instructions with stack memory references, and (3) obfuscated stack strings using xor operations. These constructed strings, especially when implemented in TLS callbacks, could contain potentially malicious commands, API names, or URLs that are only visible during runtime, making detection difficult for security tools. By identifying various implementation methods (push, mov, and xor), the detection is more comprehensive than relying on a single instruction pattern.

*5.3.4. API Hashing Techniques*

This is a stealthy method where attackers use mathematical operations (like xor/rol/ror) followed by indirect calls to obscure which Windows APIs they're accessing. The code specifically looks for sequences of these operations followed by function pointer de-referencing. When used in TLS callbacks, this technique helps malware hide its intended functionality and evade detection by concealing its API calls.

*5.3.5. NOP-Sled Attack*

NOP sled detection in TLS callbacks is critical for security analysis as these callback functions execute with the same privileges as the main program but receive less scrutiny. Malware authors implement sequences of NOP instructions within TLS callbacks to create enlarged target areas for buffer overflow attacks [15], increasing the probability of successfully redirecting execution flow to malicious payloads despite addressing inaccuracies. Since TLS callbacks execute early in the process lifetime, these buffer overflow techniques can compromise systems before standard security controls engage, making their detection an essential indicator of potential memory corruption exploitation attempts.

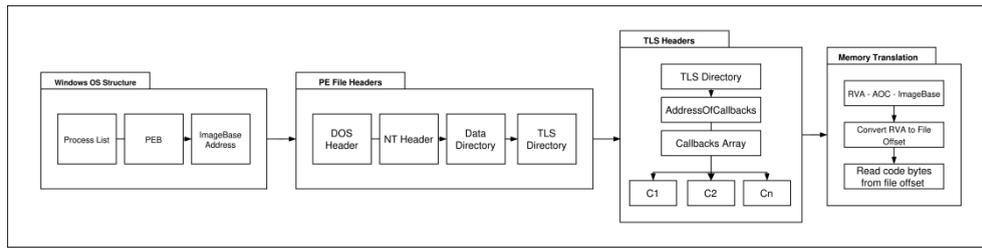

**Fig. 3.** Illustration of the TLS Callbacks extraction process.

*5.4. Apparatus and Test Cases*

The setup for this project consisted of a Linux machine running Volatility 3, the most recent version of the memory forensics framework. A custom plugin was developed and tested using memory samples obtained from a Windows 10 system. To create specific scenarios for testing, custom implementations and samples were generated and executed on a Windows 10 machine.

After the execution was completed, a memory dump was captured using VMware's snapshot feature, which produced a .vmem file for testing the plugin. The memory sample was obtained from a machine equipped with 2GB of RAM and an Intel i5 processor.

6. **Methodology**

Figure 3 illustrates a detailed representation of the working of our plugin. In short, the methodology for analyzing TLS callback-based threats consists of six main phases.

- **Memory Structure Analysis:** The plugin locates and extracts TLS-related data structures from Windows process memory, supporting both 32-bit and 64-bit architectures. This involves systematic parsing of PE headers and TLS directories to identify callback locations.
- **Pattern Detection Implementation:** A detection system was developed using the Capstone disassembly framework. This system analyzes instruction patterns, code context, and behavioral characteristics to identify potentially malicious TLS callback implementations.
- **Multi-layered Analysis:** The plugin employs multiple detection mechanisms including instruction pattern matching, YARA rule integration, and memory access analysis. These layers work together to provide thorough coverage of various attack techniques that abuse TLS callbacks.
- **API Resolutions:** The plugin identifies dynamically linked API functions by examining the Import Address Table (IAT), which contains the addresses of imported functions. By matching the target addresses of call or jump instructions with entries in the IAT, the plugin can determine the corresponding API names, even when functions are referenced indirectly.
- **Recursive Disassembly:** When unknown call instructions to memory addresses are identified, the plugin implements a recursive disassembly capability that follows these execution paths, disassembling the target addresses to reveal potentially hidden suspicious functionality that might otherwise remain undetected in standard analysis (see Appendix A).
- **Result Presentation:** The findings are presented through a structured output format that includes process details, memory locations, disassembled code, and identified suspicious patterns (if the flag is mentioned), enabling efficient analysis of potential threats.

| Algorithm 1: PLUGIN TLSCHECK: |
|---|
| **INPUT:** |
|    - Memory Dump |
|    - Optional: Process IDs, disassembly bytes, custom regex, YARA rules |
| **For each process in memory dump:** |
|    - Extract executable file |
|    - Analyze PE file structure |
|    - Locate *IMAGE_TLS_DIRECTORY* structure |
|     If *IMAGE_TLS_DIRECTORY* exists: |
|      *If 32-bit or 64-bit executable:* |
|       - Extract TLS Callback address |
|       - Disassemble TLS callback instructions |
|       - Disassemble unknown calls |
|       - Perform API resolution |
|      *Perform Instruction Analysis:* |
|       - Detect suspicious instructions |
|       - Match against custom regex |
|       - Scan with optional YARA rules |
|      *Output:* |
|       - Process details |
|       - TLS callback information |
|       - Suspicious instruction findings |
|       - YARA rule matches (if applicable) |
|     End |
|    End |
| End |

## 6.1. Extraction of callbacks

The TLS callback extraction and analysis process begins by interfacing with Volatility 3's memory analysis framework to identify and extract running processes from memory dumps. For each process, the plugin follows a systematic approach to extract, parse, and analyze TLS callbacks (see algorithm 1). Initially, the plugin dumps the complete process from memory using the Process Environment Block (PEB) to locate the process base address. The process image is reconstructed by parsing the `IMAGE_DOS_HEADER` structure at the base address,

```
kali@kali:~/Tools/volatility3$ python3 vol.py -f ../../TLS/TLS.vmem windows.TlsCheck --pid 4268
Volatility 3 Framework 2.11.0
Progress:  100.00               PDB scanning finished
PID     PPID    Process Name    Offset(V)       TLS RVA(V)      Architecture    Path

4268    1552    TLS.exe 0x8c85a4a77080  0x2180  x86     C:\Users\Flare\Desktop\Samples\Legit\TLS.exe

--------------------------------------------------------------
TLS-Callback Found in Process: TLS.exe (PID: 4268)
Address range: 0x401000 - 0x401040
--------------------------------------------------------------
64 a1 2c 00 00 00 6a 00 68 10 21 40 00 68 10 21  d.,...j.h.!@.h.!
40 00 8b 00 6a 00 c7 40 04 1f 02 00 00 ff 15 98  @...j..@........
20 40 00 c2 0c 00 cc cc cc cc cc cc cc cc cc cc   @..............
64 a1 2c 00 00 00 8b 00 ff 70 04 68 20 21 40 00  d.,......p.h !@.
Disassembly:
0x401000:       mov     eax, dword ptr fs:[0x2c]
0x401006:       push    0
0x401008:       push    0x402110
0x40100d:       push    0x402110
0x401012:       mov     eax, dword ptr [eax]
0x401014:       push    0
0x401016:       mov     dword ptr [eax + 4], 0x21f
0x40101d:       call    dword ptr [0x402098]
0x401023:       ret     0xc
--------------------------------------------------------------
```

**Fig. 4.** General output representation of our plugin.

```
kali@kali:~/Tools/volatility3$ python3 vol.py -f ../../TLS/TLS.vmem windows.TlsCheck --help
Volatility 3 Framework 2.11.0
usage: volatility windows.TlsCheck.TLSCheck [-h] [--pid [PID ...]] [--disasm-bytes DISASM-BYTES] [--scan-suspicious]

options:
  -h, --help            show this help message and exit
  --pid [PID ...]       Process IDs to include (all other processes are excluded)
  --disasm-bytes DISASM-BYTES
                        Bytes to disassemble (Default: 64)
  --scan-suspicious     Displays suspicious TLS Callback instruction(s) along with the disassembly
  --regex REGEX         Custom regex pattern to match against disassembled instructions
  --yara-file YARA-FILE
                        Path to custom YARA rule file
```

**Fig. 5.** Options provided by *TlsCheck*.

followed by the `IMAGE_NT_HEADERS`. The dumped process is then analyzed using the PE format structure, where the plugin specifically targets the Data Directory entries to locate the TLS Directory (located at index 9). The TLS Directory's location is calculated by converting the Relative Virtual Address (RVA) to a file offset using section headers, where each section's `VirtualAddress` and `PointerToRawData` values are used to compute the correct file position.

For 32-bit processes, the plugin parses the IMAGE_TLS_DIRECTORY structure, which contains crucial fields including `StartAddressOfRawData`, `EndAddressOfRawData`, `AddressOfIndex`, `AddressOfCallBacks`, `SizeOfZeroFill`, and Characteristics. The `AddressOfCallBacks` field, a critical component, points to an array of function pointers. The plugin calculates the actual callback addresses by subtracting the image base from the virtual addresses to obtain RVAs. These RVAs are then converted to file offsets using section header information for precise location of the callback code in the dumped process image.

For 64-bit processes, the `IMAGE_TLS_DIRECTORY` structure is parsed similarly, but with appropriate 64-bit pointer handling. The callback array traversal continues until a null pointer is encountered, indicating the end of the callback array. Each discovered callback address undergoes architecture-specific address translation (using the image base and section mappings) to locate the actual callback code in the dumped process image.

When converting virtual addresses to file offsets, the plugin carefully validates section boundaries and handles cases where addresses might fall between sections. The section headers are parsed to find the most appropriate section containing the target address, using both `VirtualAddress` and `PointerToRawData` fields to compute correct file offsets.

For disassembly purposes, the plugin reads a configurable number of bytes (default: 64 bytes) from each callback location. These bytes are then processed using the Capstone disassembly engine, with architecture-specific configurations (CS_ARCH_X86 with either CS_MODE_32 or CS_MODE_64). The disassembled instructions undergo multiple static analysis phases focusing on both syntactic pattern matching and semantic context evaluation. Our analysis pipeline identifies potentially suspicious constructs (including API hashing, stack string construction, and NOP sleds (detailed in section 5.3) while examining execution context to differentiate legitimate code from malicious techniques. This combined approach evaluates instructions against multiple detection heuristics, prioritizing results based on security implications. Figure 4 represents a general output of our plugin.

## 6.2. Options Provided

We have provided five optional arguments to be passed to TlsCheck plugin for enhancing its functionality (see figure 5).
- **--pid:** Allows analysts to specify particular Process IDs for targeted investigation, enabling focused

analysis of specific processes while excluding others from the scan.
- **--disasm-bytes:** Controls the number of bytes to disassemble when examining TLS callbacks, defaulting to 64 bytes.
- **--scan-suspicious:** Enables detection and highlighting of potentially malicious instruction patterns within TLS callbacks. When enabled, it employs advanced heuristics to identify suspicious code constructs and patterns commonly associated with malicious behavior.
- **--regex:** Accepts user-defined regular expression patterns to match against disassembled instructions.
- **--yara-file:** Enables integration with custom YARA rules, allowing analysts to leverage their existing YARA ruleset for identifying known malicious

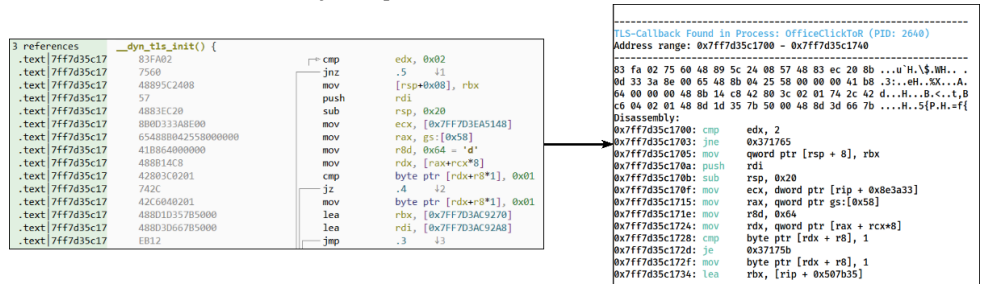

**Fig. 6.** Correlation of disassembly between Malcat and TlsCheck

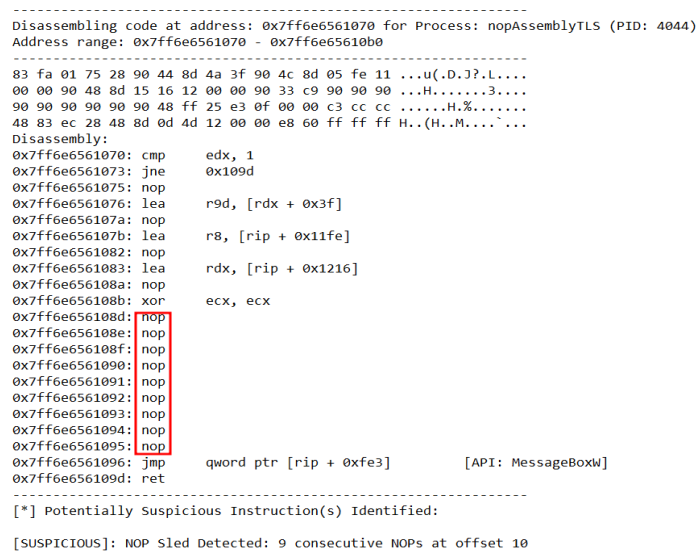

**Fig. 7.** Suspicious NOP sleds detected by *TlsCheck*

patterns within TLS callbacks.

### 6.3. Working of the plugin

The TlsCheck plugin operates by analyzing Windows memory dumps to identify and extract TLS callbacks from running processes. Upon execution, the plugin scans the memory dump to enumerate all processes and locates their TLS directory structures. It then extracts the Relative Virtual Address (RVA) of the TLS callbacks and lists processes where the RVA is not zero, indicating the presence of TLS callbacks.

To enhance the accuracy of its results, the plugin employs a filtering mechanism that automatically excludes processes with an RVA value of zero, minimizing noise and ensuring that analysts focus on relevant data. If an identified TLS callback contains an unresolved API call—meaning the address does not map to a known function—the plugin attempts to disassemble the unknown call to provide further insights (see figure 11 - Appendix). The disassembly output was cross-validated against the Malcat tool to confirm its accuracy (see Figure 6).

We also tested the plugin on a memory dump of the Ursnif malware. However, we were not able to extract TLS callbacks from the sample. To further investigate, we loaded the Ursnif process dump into IDA Freeware, but encountered similar issues, with no callbacks resolvable from the TLS directory (see Figures 9 and 10).

Additionally, the plugin supports an optional --scan-suspicious flag, which enables additional pattern recognition techniques to identify potential anomalies within TLS callbacks. We demonstrated this scanning approach on two custom samples—one employing NOP sleds in an uncommon region and another utilizing anti-debugging checks within the TLS callback routine (see Figures 7 and 8). This mode detects suspicious instruction sequences, which are commonly linked to exploit attempts and code obfuscation.

By integrating these features, the TlsCheck plugin provides an efficient method for detecting and analyzing TLS callbacks, aiding in the identification of suspicious behavior within Windows memory dumps.

[figure: disassembly listing showing anti-debugging instructions]

**Fig. 8.** Suspicious anti-debugging and control flow instructions detected by *TlsCheck*

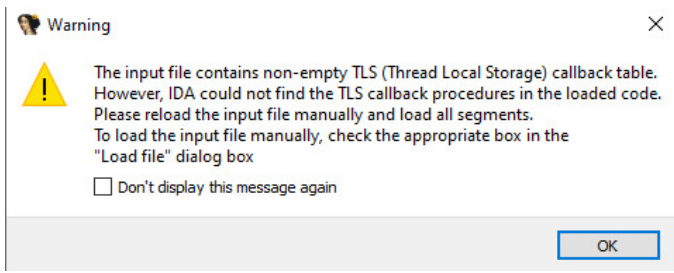

**Fig. 9.** TlsCheck output for the Ursnif variant.

[figure: IDA Freeware warning dialog about non-empty TLS callback table]

**Fig. 10.** IDA Freeware output for the Ursnif variant.

## 7. Limitations and Future Scope

While the *TlsCheck* plugin demonstrates effectiveness in analyzing TLS callbacks from memory dumps, several limitations warrant acknowledgment and present opportunities for future enhancement. A fundamental constraint of the plugin is its dependency on process presence within memory—it can only analyze TLS callbacks from processes that are actively loaded in the memory dump. This limitation impacts the plugin's ability to conduct retrospective analysis of terminated processes or examine TLS callbacks from executables that were not running during memory acquisition.

Another limitation is the scope of detection mechanisms currently implemented. While the plugin identifies TLS callbacks and flags certain suspicious behaviors, expanding its rule sets and signatures would improve its ability to detect more complex evasion techniques. Future enhancements could include analyzing TLS callbacks from hidden or unlinked processes, improving detection mechanisms for stealthy malware implementations, and incorporating sophisticated heuristics for anomaly detection. Additionally, implementing support for custom detection rules and enhanced pattern matching would allow for more flexible and thorough analysis of TLS callback behaviors. These improvements would significantly enhance the plugin's utility in forensic analysis and malware detection, particularly in cases involving advanced obfuscation and evasion techniques.

## 8. Conclusion

In this paper, we introduced a novel Volatility 3 plugin designed for the automated detection and analysis of Thread Local Storage (TLS) callbacks in memory dumps. Our plugin addresses a critical gap in existing memory analysis tools by offering comprehensive support for TLS callback structures across both 32-bit and 64-bit Windows architectures. Using a multi-layered approach that combines instruction pattern matching, YARA rule integration, and suspicious behavior detection, we provide investigators with powerful capabilities to identify potential TLS-based threats.

We demonstrated the effectiveness of our plugin through testing on diverse Windows memory dumps, successfully detecting both legitimate TLS implementations and potentially suspicious patterns. By integrating detection mechanisms for attack vectors such as code injection, control flow hijacking, and API hashing techniques, our solution delivers practical value for malware analysis and incident response scenarios.

While we acknowledge current limitations, such as the dependency on process presence, these areas present opportunities for further research. Our work represents a significant step forward in memory forensics, addressing the growing need for advanced tools to counter increasingly sophisticated threats. Future efforts will focus on expanding detection capabilities, enhancing API resolution, and developing more advanced heuristics to identify emerging TLS-based threats.

By contributing this plugin to the open-source Volatility community, we aim to strengthen collective memory forensics capabilities. Our modular and extensible design lays the groundwork for continued innovation and collaboration, enabling the forensics community to adapt to evolving cyber threats effectively.

[figure: disassembly showing unknown API calls and NOP sled detection]

**Fig. 11.** Disassembling of an unknown function call.

## Acknowledgements

We would like to thank Harsh Upadhyay for helping us create small programs and memory dumps that were used to test our plugin. His support throughout the development phase was consistent and helpful, making the testing process smoother and more manageable.

## Appendix A

**MD5 of Ursnif malware variant employing TLS Callback abuse:** 13794D1D8E87C69119237256EF068043
The plugin: https://github.com/KartikIyerr/TlsCheck

The output of an unknown call disassembly found within a TLS Callback Context.